\documentclass[english,superscriptaddress,nofootinbib]{revtex4-2}
\usepackage[T1]{fontenc}
\usepackage[latin9]{inputenc}
\usepackage{amsmath}
\usepackage{graphicx}
\usepackage{babel,comment}
\usepackage{color,diagbox}
\usepackage{array}
\usepackage{subfigure}
\usepackage{dutchcal}
\usepackage{ulem}
\graphicspath{{pic/}}
\begin{document}
\title{Deconfinement and chiral restoration phase transition under rotation from holography in an anisotropic gravitational background}

\author{Yidian Chen}
\email[]{chenyidian@hznu.edu.cn}
\affiliation{School of Physics, Hangzhou Normal University, Hangzhou, 311121, P.R. China}

\author{Xun Chen}
\email[]{chenxun@usc.edu.cn}
\affiliation{School of Nuclear Science and Technology, University of South China, Hengyang 421001, China}

\author{Danning Li}
\email[]{lidanning@jnu.edu.cn}
\affiliation{Department of Physics and Siyuan Laboratory, Jinan University, Guangzhou 510632, P.R. China} 

\author{Mei Huang}
\email[]{huangmei@ucas.ac.cn}
\affiliation{School of Nuclear Science and Technology, University of Chinese Academy of Sciences, Beijing 100049, China}	

\begin{abstract}
We investigate the effects of rotation on deconfinement and chiral phase transitions in the framework of dynamical holographic QCD model. Instead of transforming to the rotating system by Lorentz boost, we construct an anisotropic gravitational background by incorporating the rotating boundary current. We firstly investigate the pure gluon system under rotation to extract deconfinement phase transition from the Polyakov loop then add 2-flavor probe for chiral restoration phase transition from the chiral condensate. It is observed that at low chemical potentials, the deconfinement phase transition of pure gluon system is of first order and the chiral phase transition of 2-flavor system is of crossover. Both the critical temperatures of deconfinement and chiral phase transitions decrease/increase with imaginary/real angular velocity ($\Omega_I/\Omega$) as $T/T_c\sim 1- C_2 \Omega_I^2$ and $T/T_c\sim 1+ C_2 \Omega^2$, which is consistent with lattice QCD results. In the temperature-chemical potential $T-\mu$ phase diagram, the critical end point (CEP) moves towards regions of higher temperature and chemical potential with real angular velocity.
\end{abstract}
\maketitle

\section{Introduction}

In non-central ultra-relativistic heavy ion collisions, the strength of magnetic field can reach the magnitude of $eB\sim10^{18-20}\ \text{Gauss}$  \cite{Skokov:2009qp,Deng:2012pc}  and the angular momentum can reach the strength of  $\mathcal{O}(\text{10}^4-\text{10}^5) \hbar $ \cite{Becattini:2007sr} with local angular velocity in the range of $0.01\sim 0.1$ GeV and the angular velocity of order $\omega \sim 10^{21}\text{s}^{-1}$ \cite{Li:2017slc}. QCD matter under strong magnetic field and fast rotation has attracted lots of interests in recent decades \cite{Kharzeev:2007jp, Kharzeev:2010gr,Bali:2012zg, Andersen:2014xxa, Miransky:2015ava,Hattori:2023egw,Liang:2004ph,Liang:2004xn}.

It was predicted that the global spin polarization for the final state hyperons \cite{Liang:2004ph} as well as the vector meson spin alignment \cite{Liang:2004xn} would occur due to the spin-orbit coupling of quantum chromodynamics (QCD). The global polarization of $\Lambda$ and $\bar{\Lambda}$ hyperons was observed in non-central heavy-ion collisions in 2017 \cite{STAR:2017ckg,STAR:2018gyt,STAR:2020xbm}. The spin alignment was reported  with $\rho_{00}<1/3$ by the ALICE Collaboration for both $\phi$ and $K^{*0}$ vector mesons at small transverse momenta \cite{ALICE:2019aid} in 2020, while STAR Collaboration gave $\rho_{00}>1/3$ for $\phi$ meson and $\rho_{00}\approx 1/3$ for $K^{*0}$ \cite{STAR:2022fan} in 2023. The inconsistent results attracted lots of discussions  \cite{Sheng:2019kmk,PhysRevD.105.099903, Sheng:2020ghv,Xia:2020tyd,Muller:2021hpe, Yang:2021fea,Goncalves:2021ziy,Li:2022neh, Wagner:2022gza,Wagner:2022gza,Kumar:2022ylt,Sheng:2022wsy,Sheng:2022ffb,Kumar:2023ghs,Kumar:2023ojl}.

QCD matter and the chiral restoration and deconfinement phase transitions under rotation have also attracted much attention in recent years. It was found in effective QCD models that the critical temperature of chiral phase transition is suppressed by the rotation in Refs. \cite{Chen:2015hfc,Jiang:2016wvv,Ebihara:2016fwa, Chernodub:2016kxh,Chernodub:2017ref,Wang:2018sur,Sun:2021hxo, Xu:2022hql,Sun:2023kuu}. In the framework of holographic QCD in Ref. \cite{Chen:2020ath}, it was found that the critical temperature of deconfinement phase transition decreases with angular velocity, which is confirmed by other holographic studies \cite{Braga:2023qej,Li:2023mpv, Ambrus:2023bid,Zhao:2022uxc,Golubtsova:2022ldm, Chen:2022mhf,Chen:2022smf,Yadav:2022qcl,Braga:2022yfe, Cartwright:2021qpp,Golubtsova:2021agl}. However, lattice QCD simulations \cite{Braguta:2021jgn,Braguta:2022str,Yang:2023vsw} showed that the critical temperature of the deconfinement phase transition in gluon dynamics as well as the chiral restoration phase transition in 2+1 flavor system increases with angular velocity, which is opposite to above results obtained in effective and holographic models. 

The contradiction between the effective and holographic models calculations and lattice results on rotation attracted much attention and comprehensive analysis \cite{Chernodub:2020qah,Fujimoto:2021xix,Cao:2023olg,Jiang:2023zzu,Chen:2023cjt}. Recently in the Polarized-Polyakov-loop Nambu-Jona-Lasinio (PPNJL) model shown in Ref. \cite{Sun:2024anu}, which takes into account the rotating gluon background, it can be realized that both critical temperatures of deconfinement and chiral phase transition increase with the angular velocity in consistent with lattice results. This confirms that the gluodynamics is sensitive to rotation as implied from the Zeeman-like behavior of spin-1 vector meson under rotation observed in \cite{WeiMingHua:2020eee}.

The lesson learnt from Ref. \cite{Sun:2024anu} is that the rotating gluon background has to be properly taken into account. Therefore, in this work we reanalyze the rotating gluodynamics in the framework of holographic QCD model then add flavor probes on this background. In previous holographic studies, many efforts have been made in understanding the rotating effect on QCD phase transitions, besides studies in the Kerr black hole background \cite{McInnes:2014haa,Erices:2017izj,Mcinnes:2018xxz,McInnes:2018mwj,Arefeva:2020jvo}. One common way to introduce angular velocity is to do a simple coordinate transformation to the dual gravity backgrounds \cite{Awad:2002cz,Sheykhi:2010pya,BravoGaete:2017dso,Nadi:2019bqu,Chen:2020ath}, which have been carefully tested in non-rotational systems. In this approximation, the rotating effect can be partially encoded in the holographic approach. It is shown that the transition temperature of deconfining phase transiton decreases with increasing angular velocity, as discussed above. But the effect on the scalar operators, like the chiral condensation, which is the order parameter of chiral phase transtion, has not been studied. The reason is that the scalar operator is invariant under the coordinate transformation.

In Ref. \cite{Chen:2022mhf}, an alternative way was applied to consider the effects on the chiral phase transtion. By introducing rotating boundary currents and solving the soft-wall model in the probe limit (as also in Refs. \cite{Domenech:2010nf,Keranen:2009ss,Keranen:2009re,Dias:2013bwa,Li:2019swh}), the rotational effect on chiral condensation can be obtained. It gives a similar effect with that on the deconfining transition, i.e. decreasing transition temperatures with increasing angular velocities. However, since the analysis is done in the probe limit, the corrections from the rotating boundary currents can not be encoded in the gravity background, and thus the effects on the gluodynamics have not been incorporated in this way. Obviously, if one tries to go beyond the probe limit, it is possible to get the rotational effects on both gluo-dynamics and chiral dynamics simultaneously. However, to calculate the full gravity background dual to a rotating system is not an easy task. Hence, in this work, we will focus on the near-centre region of the rotating system, in which we can take a simple ansatz on the gravity background for the rotating system. After solving the system, we will try to extract the rotating effects on the dynamics of both the two sectors.

The organization of the paper is as follows. In Sec. \ref{sec:model}, the dynamical holographic QCD model is reviewed and extended to the anisotropic case. Thermodynamic quantities under rotation are explored in Sec. \ref{sec:thermal}. Deconfinement and chiral phase transitions with angular velocity are studied in Sec. \ref{sec:confinement} and \ref{sec:chiral}, respectively. Finally, a summary and discussion are provided in Sec. \ref{sec:sum}.

\section{Model setup}
\label{sec:model}

The holographic QCD method, based on the Anti-de Sitter/conformal field theory (AdS/CFT) correspondence~\cite{Maldacena:1997re,Gubser:1998bc,Witten:1998qj}, has become one of the important nonperturbative methods to understand the properties of QCD and its phase transitions. In the past, the holographic method has been successfully applied in various aspects, including transport coefficients \cite{Policastro:2001yc,Buchel:2003tz,Kovtun:2004de}, thermodynamical quantities \cite{Gubser:2008ny,Gursoy:2007er,Gursoy:2007cb,Grefa:2021qvt,Cai:2022omk,Li:2011hp}, hadronic physics \cite{Erlich:2005qh,Karch:2006pv,deTeramond:2005su}, and so on. Building on these studies, the dynamical holographic QCD model (DHQCD) \cite{Li:2013oda} has further simulated the dynamical behavior of QCD and successfully described the confinement/deconfinement and chiral phase transitions \cite{Li:2014hja,Li:2014dsa,Fang:2015ytf,Chen:2019rez}. However, to fully understand the behavior of QCD under extreme conditions, especially the impact of rotational effects, it is necessary to extend the DHQCD model. In this section, we review the DHQCD model and present a new extension to the model which allows it to be applied to the rotating case.

\subsection{The Einstein-Maxwell-Dilaton system}

From Ref. \cite{Chen:2019rez}, the quenched DHQCD model with chemical potential is described by the Einstein-Maxwell-dilaton system. The action in the string frame can be written as
\begin{eqnarray}
    S_{\rm tot}^s  &=& S_G^s+S_M^s, \label{equ:action}\\
    S_G^s &=&\frac{1}{16\pi G_5}\int d^5x\sqrt{-g^s}e^{-2 \Phi }\left[R^s+4\partial_M \Phi \partial ^M \Phi - V^s(\Phi)-\frac{h(\Phi)}{4}e^{\frac{4\Phi}{3}}F_{MN}F^{MN}\right], \label{equ:background}\\
    S_M^s &=&-\int d^5x
    \sqrt{-g^s}e^{-\Phi}{\rm Tr}\left[\nabla_{M}X^{\dagger} \nabla^{M}X+V_X(|X|,F_{MN}F^{MN})\right]\label{equ:matter},
\end{eqnarray}
with lowercase $s$ for string frame and 5D Newton constant $G_5$. In the DHQCD model, the total action $S_{\rm tot}^s$ can be decomposed into $S_G^s$ describing the gluo-dynamics and $S_M^s$ representing the quark dynamics. The $\Phi$ and $V^s(\Phi)$ represent the dilaton field and its potential. The strength tensor of the gauge field $A_M$ dual to the baryon number current is denoted by $F$. The coupling between the dilaton and the gauge field is given by $h(\Phi)$. In the presence of a non-zero baryon chemical potential ($\mu_B\neq 0$), the only non-trivial component of the $U(1)$ gauge field is $A_t$, i.e. $A=A_tdt$. The scalar field $X$ is associated with the quark condensate $\langle\bar{q}q\rangle$, and $V_X$ represents its potential, which is coupled to the gauge field strength. The EMD system describes the gluon dynamics quite well and enables to obtain the QCD phase transition \cite{Li:2014hja,Chelabi:2015cwn,Chelabi:2015gpc} and the glueball spectra \cite{Li:2013oda} in the pure gluon system.

The equations of motion and thermodynamics are more conveniently solved in the Einstein frame, so the background action is rewritten as
\begin{eqnarray}
    S_G^e =\frac{1}{16\pi G_5}\int d^5x\sqrt{-g^e}\left[R^e-\frac{1}{2}\partial_M \Phi \partial ^M \Phi - V^e(\Phi)-\frac{h(\Phi)}{4}F_{MN}F^{MN}\right],
\end{eqnarray}
with lowercase $e$ for the Einstein frame. Here, the metric and potential satisfy $g^s_{MN}=e^{\frac{4\Phi}{3}}g^e_{MN}$ and $V^e=e^{\frac{4\Phi}{3}}V^s$, respectively.
In Einstein's framework, the metric can be written as
\begin{eqnarray}
    ds^2=\frac{L^2e^{2A_e(z)}}{z^2}\bigg[-f(z)dt^2+\frac{dz^2}{f(z)}+d\vec{x}^2)\bigg],
\end{eqnarray}
where the warp factors $A_e(z)$ and $A_s(z)$ of the two frames satisfy $A_e(z)=A_s(z)-\frac{2}{3}\Phi(z)$. Without loss of generality, the AdS radius is set to $L=1$.

There are generally two ways to solve the equations of motion. The first is through fixing $h(\Phi)$ and the explicit form of the potential $V(\Phi)$. The second, known as the ``potential reconstruction method'', is to resolve the dilaton potential by inputting function $\Phi$ or $A_e$ and $h(\Phi)$. It should be emphasized that the coefficients of the potential $V(\Phi)$ of this method are temperature and chemical potential dependent, which is in a sense consistent with the spirit of finite temperature field theory. This method is easier to handle and is a good approximation to the fixed potential method and is employed in this paper. As in Ref. \cite{Li:2014hja}, the dilaton field is required to have the following form
\begin{eqnarray}
    \Phi(z)=\mu_G^2z^2\tanh(\mu_{G^2}^4z^2/\mu_G^2).
\end{eqnarray}
In the UV region, the dilaton field has $\Phi\stackrel{z\to0}{\longrightarrow}\mu_{G^2}^4z^4$, which is dual to the gauge-invariant gluon condensation ${\rm Tr}\langle G^2\rangle$. In the IR region, the dilaton field has $\Phi\stackrel{z\to\infty}{\longrightarrow}\mu_{G}^2z^2$ in order to satisfy the linear Regge trajectory of mesons spectra.

As shown in the Tab. \ref{tab:para}, there are four parameters $\mu_G$, $\mu_{G^2}$, $G_5$, and $h(\Phi)$ in the model. For the first three parameters, the values are taken from Ref. \cite{Li:2014hja}. For the coupling $h(\Phi)$ of the dilaton and gauge field, it is chosen as $h=4$, in order to realize the deconfinement phase transition in the $T-\mu$ plane with critical endpoint (CEP) $(T_{CEP},\mu_{CEP})=(0.271,0.14)$GeV. The $T-\mu$ phase diagram is consistent with the deconfinement phase diagram for heavy quarks.

\renewcommand\arraystretch{1.5}
\begin{table}[!htbp]
	\centering
	\begin{tabular}{|p{2cm}<{\centering}|p{2cm}<{\centering}|p{2cm}<{\centering}|p{2cm}<{\centering}|p{2cm}<{\centering}|p{2cm}<{\centering}|}
	\hline
	Parameters & $\mu_G$ (GeV) & $\mu_{G^2}$ (GeV) & $G_5$ & $h$ & $\mu_\Omega$ (GeV$^{-1}$)\\
	\hline
	Values & 0.75 & 3 & 1.25 & 4 & 122\\
	\hline
    \end{tabular}
	\caption{\label{tab:para} The parameters of the DHQCD model, where the first three parameters were selected from Ref. \cite{Li:2014hja}. The last parameter $\mu_G$ is set as described in Sec. \ref{sec:anisotropic}.}
\end{table}

\subsection{Anisotropic background under rotation}
\label{sec:anisotropic}

To investigate the QCD phase transition under rotation, one is faced with the challenge of solving the EMD system under a rotating black hole. The complexity of the problem is that in the cylindrical coordinate system, the metric, dilaton, and gauge fields all depend on five-dimensional coordinates $z$ and the radial direction $r$. As a result, different methods have been used to study the effects of rotation.

In Ref. \cite{Chen:2020ath}, the effect of rotation on the deconfinement phase transition has been explored through the coordinate transformation. Alternatively, some have analysed inhomogeneous chiral phase transitions under the probe approximation \cite{Chen:2022mhf}. However, the results of the holographic model suggest that rotation reduces the critical temperature of the phase transition, which is inconsistent with the findings of lattice QCD \cite{Braguta:2021jgn,Yang:2023vsw}.

In order to reconstruct the results of lattice QCD, the anisotropic background is considered in this paper. As shown in Ref. \cite{Chen:2022mhf}, the scalar field, which is dual to the condensation, does not depend significantly on the radial coordinate near the center of rotation. Therefore, it can be assumed that the metric and dilaton fields are only the function of the fifth-dimensional coordinate near the center. Furthermore, an additional anisotropic function $B(z)$ needs to be taken into account in the metric since the effect of rotation. So, the background metric of the Einstein frame in the cylindrical coordinate system is
\begin{eqnarray}
    ds^{2}=\frac{L^{2}e^{2A_e(z)}}{z^{2}}[-f(z)dt^2+\frac{dz^2}{f(z)}+e^{B(z)}dr^2+r^2e^{B(z)}d\theta^2+e^{-2B(z)}dx_3^2],\label{equ:metric-rot}
\end{eqnarray}
where $r$ denotes the radial direction, $\theta$ is the direction of the polar angle, and $x_3$ is the direction of the symmetry axis. According to the holographic principle, the temperature of the system is given by the Hawking temperature, i.e.,
\begin{eqnarray}
    T=\frac{|f^\prime (z_h)|}{4\pi},
\end{eqnarray}
with the horizon $z_h$ of black brane.

As in Ref. \cite{Chen:2022mhf}, the rotation is introduced by a non-zero polar angle component $A_\theta$ of the gauge field. It can be expanded as $A_\theta\sim \Omega r^2+\rho_\theta(r,z)$. Near the center, the current $\rho_\theta$ can be neglected and the field $A_M$ is approximated as
\begin{eqnarray}\label{equ:gauge}
    A_M=(A_t,0,0,A_\theta,0),\qquad A_\theta=\Omega r^2,
\end{eqnarray}
with $A_z=0$ gauge and angular velocity $\Omega$. Although the gauge field $A_\theta$ depends on the radial coordinates $r$, the strength $F^2$ is in fact independent of $r$. Therefore, both the metric and dilaton fields depend only on the fifth-dimensional coordinate $z$. Furthermore, It should be noted that this choice naturally satisfies Maxwell's equation in the $\theta$-direction.

It is important to note that Eqs. \eqref{equ:metric-rot} and \eqref{equ:gauge} hold only under the near-centre approximation. If the system is far from the center of the rotation axis, then the radial direction dependence and the current $\rho_\theta$ are not negligible. If near the edge of the system, the boundary has a significant effect on the profile of the fields and the finite size effect becomes indispensable. Additionally, the non-trivial current $\rho_\theta$ induces a non-diagonal term $g_{\theta j}$ to the metric.

Under the near-centre approximation, i.e., Eqs. \eqref{equ:metric-rot} and \eqref{equ:gauge}, the equations of motion of the EMD system can be obtained as
\begin{subequations}
\begin{align}
    \frac{z^2 h(\Phi ) e^{-2 A_e} \left(A_t'^2+4 e^{-2 B} \Omega ^2\right)}{4 f}+\frac{3 f'}{2
   f} \left(A_e'-\frac{1}{z}\right)+\frac{e^{2 A_e} V^e(\Phi )}{2 z^2 f} &\notag\\
   -\frac{12 A_e'}{z}+6 A_e'^2-\frac{3 B'^2}{4}-\frac{2 \Phi'^2}{3}+\frac{6}{z^2} &= 0,\\
   -z^2 e^{-2 A_e} h(\Phi ) \left(A_t'^2+\frac{8}{3}  e^{-2 B} \Omega ^2\right)+3f' \left(A_e'-\frac{1}{z}\right)+f'' &= 0, \label{equ:eoms-f}\\
   \frac{B'^2}{2}+\frac{4 \Phi'^2}{9}+\frac{2 A_e'}{z}-A_e'^2+A_e'' &= 0,\\
   \frac{4 \Omega ^2 z^2 h(\Phi ) e^{-2 \left(A_e+B\right)}}{3 f}+B' \left(3 A_e'+\frac{f'}{f}-\frac{3}{z}\right)+B'' &= 0,\\
   A_t' \left(A_e'+\frac{h'}{h}-\frac{1}{z}\right)+A_t'' &= 0, \label{equ:eoms-at}\\
   \frac{z^2 e^{-2 A_e} \partial_\Phi h(\Phi ) \left(A_t'^2-4 e^{-2 B} \Omega ^2\right)}{2 f}-\frac{e^{2 A_e} \partial_\Phi V_e(\Phi)}{z^2 f}+\Phi'\left(3
   A_e'+\frac{f'}{f}-\frac{3}{z}\right)+\Phi '' &= 0,\label{equ:eoms-dilaton}
\end{align}
\end{subequations}
with the derivative $'$ with respect to $z$.

As shown in Ref. \cite{Sun:2024anu}, gluo-dynamics is sensitive to rotation, so the coefficients $\mu_G$ and $\mu_{G^2}$ of the dilaton field should depend on the angular velocity $\Omega$. In order to realize rotation-dependent gluo-dynamics, the following simple form of the dilaton field is considered
\begin{eqnarray}\label{equ:dilaton}
    \Phi=(\mu_G+\mu_\Omega\Omega^2)^2z^2\tanh(\mu_{G^2}^4z^2/(\mu_G+\mu_\Omega\Omega^2)^2).
\end{eqnarray}
In this paper, only the parameter $\mu_G$ is taken into account for the angular velocity $\Omega$ dependence. Thus a new parameter $\mu_\Omega$ is introduced in the profile of the dilaton field. In addition, the angular velocity in quadratic form here is considered for the case where it can be extended to complex plane, i.e. to imaginary angular velocity $\Omega_I=-i\Omega$. Of course more complicated forms of $\mu_G(\Omega)$ can be considered, such as $\mu_G+\mu_\Omega\Omega^2+\mu_{\Omega_2}\Omega^4+...$ . When the angular velocity is not very high $\Omega \leq 0.1$ GeV, the effect of the leading order is considered.

The appropriate boundary conditions need to be chosen in order to solve the equations of motion Eqs. (\ref{equ:eoms-f}-\ref{equ:eoms-at}). At the ultraviolet (UV) boundary, the background geometry is required to be asymptotically AdS$_5$ spacetime, so
\begin{eqnarray}
   f(0) = 1,\qquad A_e(0) = 0,\qquad B(0) = 0.
\end{eqnarray}
At the infrared (IR) boundary, $A_e$ and $B$ are required to satisfy the natural boundary conditions, while the blackening factor has $f(z_h)=0$. According to the holographic principle, the gauge field $A_t(z)$ is required to have the following boundary conditions
\begin{eqnarray}
   A_t(0)=\mu,\qquad A_t(z_h)=0.
\end{eqnarray}

\section{Thermodynamics of the pure gluon system under rotation}
\label{sec:thermal}

In the previous section, we have successfully extended the DHQCD model to anisotropic backgrounds. This extension allows us to describe rotating QCD matter. In this section, we applied the holographic method to calculate the thermodynamic quantities of the system. Specifically, we calculate the free energy, entropy and specific heat capacity, and so on. Further, using the free energy obtained, the model is able to fit the relationship between the phase transition temperature $T_c(\Omega)$ and the angular velocity $\Omega$ of lattice QCD, by tuning the new parameter $\mu_\Omega$.

In this paper, thermodynamic quantities are calculated by integral of entropy density. According to the Bekenstein-Hawking formula, the entropy density of the black brane can be obtained as
\begin{eqnarray}
   s=\frac{e^{3A_e(z)}}{4G_5z^3}|_{z=z_h},
\end{eqnarray}
with five-dimensional Newton constant $G_5$ listed in Tab. \ref{tab:para}. From thermodynamics, the total differential of Landau free energy density $\mathcal{f}$ of the giant canonical system is
\begin{eqnarray}
   d\mathcal{f}=-sdT-\rho d\mu-jd\Omega.
\end{eqnarray}
with energy density $\epsilon$, baryon number density $\rho$, and angular momentum density $j$. The free energy density can be obtained by integrating the entropy density when the chemical potential and angular velocity are fixed, which contains an integration constant. For pure glue system, as shown in Ref. \cite{He:2013qq}, $\mathcal{f}(z_h\to\infty)=0$ is required to satisfy, then the free energy density can be written as
\begin{eqnarray}
   \mathcal{f}=\int_{z_h}^\infty s\frac{dT}{dz_h}dz_h.
\end{eqnarray}
For the pressure $P$ and energy density $\epsilon$, they can be given by the following relation
\begin{eqnarray}
   P=-\mathcal{f}, \qquad \epsilon=\mathcal{f}+Ts+\mu\rho+\Omega j.
\end{eqnarray}
In addition, the speed of sound $C_s$ and latent heat $C_V$ of the system can also show messages about phase transition. Both these thermodynamic quantities can be obtained from above quantities as
\begin{eqnarray}
   C_s^2=\frac{dP}{d\epsilon}\bigg|_{\mu,\Omega},\qquad C_V=\frac{d\epsilon}{dT}\bigg|_{\mu,\Omega}.
\end{eqnarray}
Besides the method employed in this paper, another approach to obtain the free energy density and other thermodynamic quantities is through the renormalization on-shell action. In principle, the two approaches are equivalent.

\begin{figure}[!htbp]
	\centering
	\includegraphics[width=0.48\textwidth]{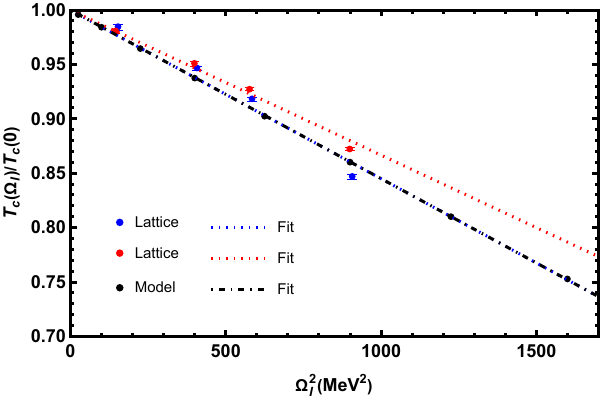}
	\caption{\label{fig:fit} The phase transition temperature as a function of the imaginary angular velocity. The $T_c(0)$ represents the phase transition temperature without rotation. The red and blue points in the figure show the lattice QCD results from Ref. \cite{Braguta:2021jgn}, and the black points show the calculations for the model. The dashed lines indicate the quadratic fitting curves.}
\end{figure}

With the free energy density given above, the phase transition temperature $T_c$ can be obtained for different angular velocities. The only parameter $\mu_\Omega$ to be determined in the DHQCD model is based on the relationship between the phase transition temperature $T_c(\Omega_I)$ and the imaginary angular velocity $\Omega_I$ predicted by lattice QCD \cite{Braguta:2021jgn}. Specifically, this relationship is given by a quadratic function
\begin{eqnarray}
   T_c(\Omega_I)/T_c(0)=1-C_2\Omega_I^2
\end{eqnarray}
with $C_2\simeq 150 \text{GeV}^{-2}$ \cite{Braguta:2021jgn}. Here $T_c(0)$ represents the phase transition temperature without rotation. In Ref. \cite{Braguta:2021jgn}, three different boundary conditions are considered and similar results are observed. Without loss of generality, the result from periodic boundary condition is chosen to fit the parameter in this paper. The results of fitting are displayed in Fig. \ref{fig:fit}. The black points in the figure are the model's results, the red and blue points are the results of lattice QCD, and all the dashed lines indicate the quadratic fitting curves. It should be noted that the red and blue points correspond to different lattice size. Since the infinite volume limit cannot be chosen for the rotating system, i.e., the finite volume effect is not negligible, the relationship between the transition temperature and imaginary angular velocity depends on the size. In the DHQCD model, the near-center limit is taken into account, so such phenomena found in lattice QCD cannot be observed. However, in Ref. \cite{Chen:2022mhf}, the holographic model indeed found the dependence of the phase transition on the linear velocity. In the study, blue dots have been chosen to fit the model's parameter to enhance the effect of rotation on the phase transition. The value of the parameter $\mu_\Omega$ is shown in Tab. \ref{tab:para}. As can be seen in Fig. \ref{fig:fit}, the results of the model agree well with the quadratic function. This is comprehensible because the effect of angular velocity in the dilaton field is the quadratic.

\begin{figure}[!htbp]
    \centering
    \subfigure[entropy density]{\label{subfig:sT3}
    \includegraphics[width=0.48\textwidth]{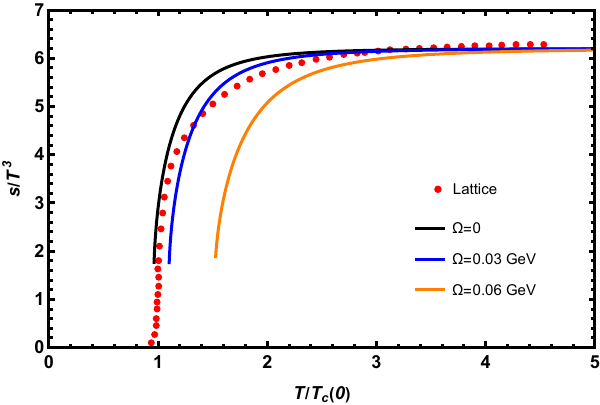}}
    \subfigure[pressure]{\label{subfig:pT4}
    \includegraphics[width=0.48\textwidth]{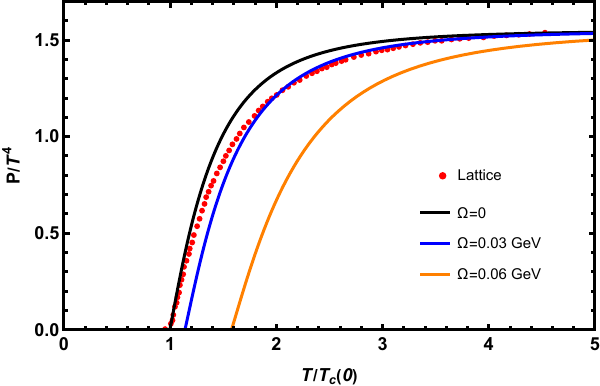}}
    \subfigure[energy density]{\label{subfig:eT4}
    \includegraphics[width=0.48\textwidth]{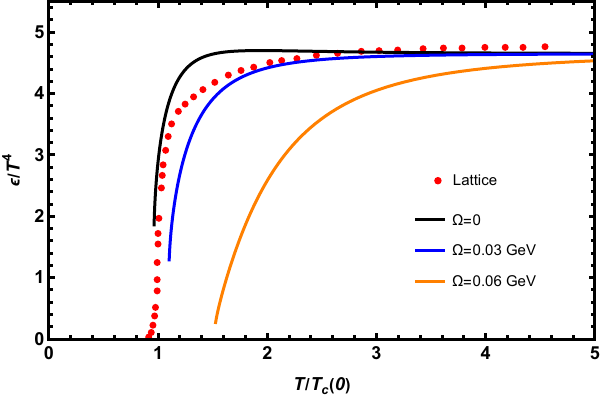}}
    \subfigure[trace anomaly]{\label{subfig:e3pT4}
    \includegraphics[width=0.48\textwidth]{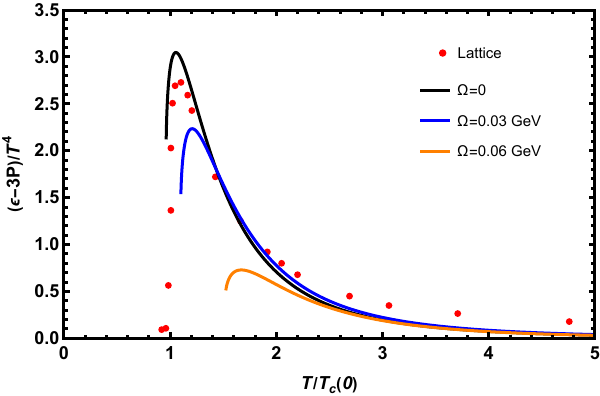}}
    \caption{\label{fig:thermal1} The entropy density, pressure, energy density and trace anomaly under rotation. The red dots in the figure are lattice QCD results from Ref. \cite{Boyd:1996bx}.}
\end{figure}

\begin{figure}[!htbp]
    \centering
    \subfigure[sound velocity]{\label{subfig:cs2}
    \includegraphics[width=0.48\textwidth]{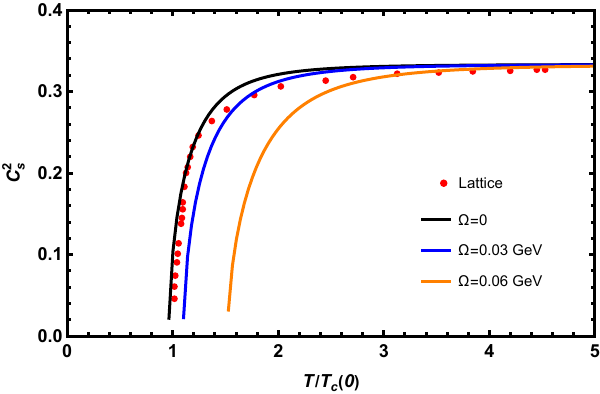}}
    \subfigure[specific heat]{\label{subfig:cv}
    \includegraphics[width=0.48\textwidth]{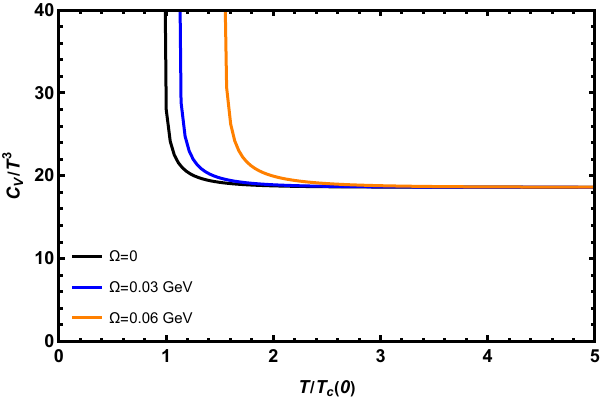}}
    \caption{\label{fig:thermal2} The square of the speed of sound and latent heat under rotation. The red dots in the figure are lattice QCD results from Ref. \cite{Boyd:1996bx}.}
\end{figure}

Within the DHQCD model, once the model parameter $\mu_\Omega$ is fixed, we are able to calculate thermodynamic quantities under rotating. The results are presented in Figs. \ref{fig:thermal1} and \ref{fig:thermal2}. In these figures, the angular velocity $\Omega$ is set to 0, 0.03, and 0.06 GeV, and we observe thermodynamic quantities such as entropy density, pressure, energy density, trace anomaly, square of the speed of sound, and latent heat. The red markers represent the results from lattice QCD for the non-rotating case, as referenced in \cite{Boyd:1996bx}.

From Fig. \ref{fig:thermal1}, it can be observed that in regions close to and far from the phase transition temperature, the thermodynamic quantities for the non-rotating case show good agreement with lattice QCD data. For the entropy density $s/T^3$ and pressure $P/T^4$, the effect of rotation is primarily a translation of the curves, with little change in the overall shape. At an angular velocity of 0.03 GeV, the translation of the curves is approximately 16$\%$. Furthermore, we observe that in the low-temperature region, thermodynamic quantities are more sensitive to the angular velocity, and as the temperature increases, thermodynamic quantities at different angular velocities tend to converge to a common value. This is qualitatively different from the results obtained in the coordinate transformation approach \cite{Chen:2020ath} but similar with that obtained in previous NJL calculation \cite{Wang:2018sur}. 

Further, if we extend the relationship between the phase transition temperature and the imaginary angular velocity to the real angular velocity, that is, the relationship $T_c(\Omega)/T_c(0)=1+C_2\Omega^2$. We can find that for an angular velocity of 0.03 GeV, the shift in the phase transition temperature satisfies $C_2\times 0.03^2\simeq 14\%$.  Thus, these two quantities can be approximated as follows
\begin{eqnarray}\label{equ:guess}
   \frac{s}{T^3}(T,\Omega)\simeq \frac{s}{T^3}\bigg(T(1+C_2 \Omega^2),0\bigg), \qquad \frac{P}{T^4}(T,\Omega)\simeq \frac{P}{T^4}\bigg(T(1+C_2 \Omega^2),0\bigg).
\end{eqnarray}
As can be seen in Fig. \ref{fig:thermal2}, for the square of the speed of sound and the latent heat, the behavior is almost the same as in $s/T^3$. Similarly, a approximation similar to Eq. \eqref{equ:guess} above can be written as
\begin{eqnarray}
   C_s^2(T,\Omega) \simeq C_s^2\bigg(T(1+C_2 \Omega^2),0\bigg), \qquad C_V(T,\Omega)\simeq C_V\bigg(T(1+C_2 \Omega^2),0\bigg).
\end{eqnarray}
However, for the energy density, the rotational effect causes the curve to become smoother within the temperature range of 1-3 $T_c(0)$, which is different from the observations made for the entropy density and pressure. This suggests that the rotational effect may manifest differently across various thermodynamic quantities. For the trace anomaly, we observe that as the angular velocity increases, the peak of the curve is suppressed. This trend may be related to the anisotropy of the system due to rotation.

\section{Polyakov loop and Deconfinement phase transition of the pure gluon system under rotation}
\label{sec:confinement}

In the previous section, we have utilized lattice QCD data to fix all model parameters and calculated the free energy density and other thermodynamic quantities under rotation. In this section, using the results from above, we further explore the Polyakov loop the anisotropic background and plot the $T-\Omega$ and $T-\mu$ phase diagrams of the deconfinement phase transition.

The Polyakov loop $\langle L\rangle$ is often used as the order parameter for the deconfinement phase transition. A nonzero $\langle L\rangle$ implies a deconfinement phase, while the vanishing $\langle L\rangle$ indicates confinement. In the holographic duality, the Polyakov loop is given by the Nambu-Goto action of the string. In gauge $\eta_1=t$ and $\eta_2=z$, the Nambu-Goto action on the world sheet can be written as
\begin{eqnarray}
   S_{NG}=\frac{g_p}{\pi T}\int_0^{z_h}dz\frac{e^{2A_s}}{z^2}\sqrt{1+f(z)(\vec{x}^\prime)^2},
\end{eqnarray}
with world sheet coordinate $\eta_1$ and $\eta_2$. With the on-shell action, the expectation value of the Polyakov loop is
\begin{eqnarray}
   \langle L(T)\rangle=e^{C_p-S_0^\prime},
\end{eqnarray}
where
\begin{eqnarray}
   S_0^\prime=\frac{g_p}{\pi T}\int_0^{z_h}dz(\frac{e^{2A_s}}{z^2}-\frac{1}{z^2}),
\end{eqnarray}
with normalization constant $C_p$ and string tension $g_p$. The expectation value of Polyakov loop for different real and imaginary angular velocities can be got by the above equations. Here, the constants $C_p$ and $g_p$ are fixed relying on lattice QCD data. It should be noted that these constants are not fixed by the results without rotation. For example, in Ref. \cite{Li:2011hp}, lattice data \cite{Gupta:2007ax} was used to fix the constants. However, in rotating gluodynamics, the 
infinite volume limit cannot be chosen, and therefore the expectation value of $\langle L\rangle$ in data \cite{Gupta:2007ax} and date \cite{Braguta:2021jgn} cannot be compared directly. For this reason, only lattice data \cite{Gupta:2007ax} have been chosen to fit the constants, ensuring their values are consistent for both rotation and non-rotation.

\begin{figure}[!htbp]
	\centering
	\subfigure[imaginary rotation]{\label{subfig:pol-i}
    \includegraphics[width=0.48\textwidth]{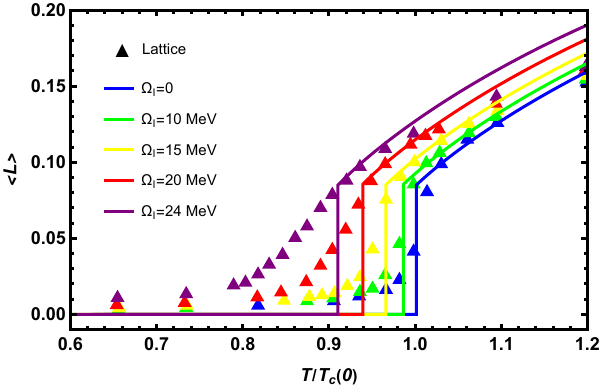}}
	\subfigure[real rotation]{\label{subfig:pol}
    \includegraphics[width=0.48\textwidth]{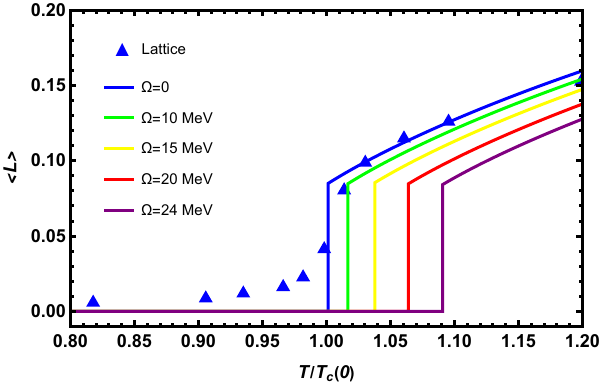}}
	\caption{\label{fig:polyakov} The expectation value of the Polyakov loop $\langle L\rangle$ obtained from the DHQCD model for pure gluon system and lattice as a function of temperature under both real and imaginary rotation. The triangles in the figure represent the lattice QCD data \cite{Gupta:2007ax} and the solid lines are the model calculations.}
\end{figure}

The results of the model calculations and lattice data under real and imaginary rotation are depicted in Fig. \ref{fig:polyakov}. The triangles represent the lattice QCD data, while the solid lines indicate the results from the model calculations. The various colors in the diagram correspond to different angular velocities, whether real or imaginary. It is important to note that the phase transition in the lattice data is not first order, that comes from finite size effects. Consequently, the data extracted from the lattice data contains information about the finite volume. The blue data points at $\Omega = \Omega_I = 0$ were used to fit the values of $C_p$ and $g_p$, resulting in $(C_p, g_p) = (1.04, 1.01)$. 

As shown in Fig. \ref{subfig:pol-i}, the model results align well with the lattice QCD data in the vicinity of the phase transition temperature. However, at a temperature of approximately $1.2 T_c(0)$, there are some discrepancies between the model and the lattice data. Within this temperature range, the results from the model calculations are more dispersed, whereas the lattice data are more clustered. This could be attributed to the effects of finite size. Additionally, it is observed that the phase transition temperatures predicted by the model differ slightly from those of the lattice QCD data. That's because the data has some deviation from relation $T_c(\Omega_I)/T_c(0)=1-C_2\Omega_I^2$. For real angular velocities, the model's predictions exhibit an inverse trend compared to those under imaginary angular velocities, as illustrated in Fig. \ref{subfig:pol}.

\begin{figure}[!htbp]
	\centering
	\subfigure[]{\label{subfig:TO}
    \includegraphics[width=0.48\textwidth]{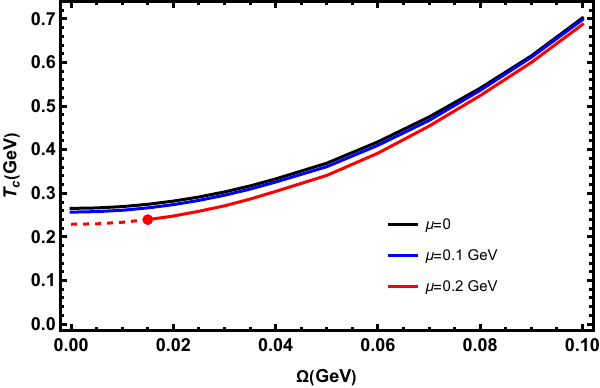}}
	\subfigure[]{\label{subfig:Tmu}
    \includegraphics[width=0.48\textwidth]{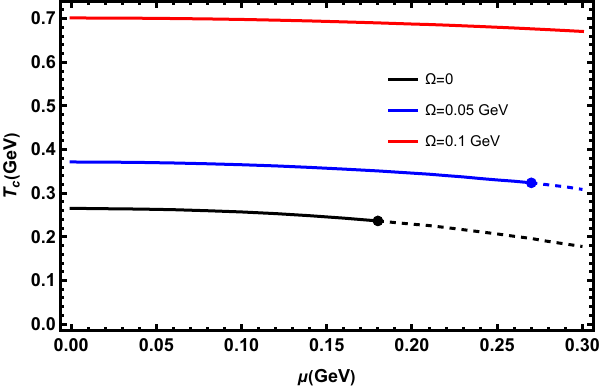}}
	\caption{\label{fig:pd} The $T-\Omega$ and $T-\mu$ phase diagrams of deconfinement phase transition for the pure gluon system.}
\end{figure}

The $T-\Omega$ and $T-\mu$ phase diagrams of the deconfinement phase transition can be plotted using the free energy densities obtained in the previous section, as shown in Fig. \ref{fig:pd}. In these diagrams, the dashed lines, solid lines, and points represent crossover, first-order phase transitions, and CEP, respectively. The pseudo-transition temperature at crossover can be defined by the second derivative, that is, $\frac{d^2L}{dT^2}=0$ or $\frac{d^2L}{d\Omega^2}=0$. 

From Fig. \ref{subfig:TO}, it can be observed that as the angular velocity $\Omega$ increases, the transition temperature $T_c$ also rises. Similar to the case of imaginary angular velocity, the relationship between the temperature $T_c$ and the angular velocity $\Omega$ approximates a quadratic function $T_c(\Omega)/T_c(0)=1+C_2\Omega^2$, which is expected. Specifically, when the angular velocity increases to 0.1 GeV, the transition temperature is approximately 2.6 to 3 times higher than that without rotation. Furthermore, the finite chemical potential can amplify the coefficient $C_2$.

As the angular velocity increases, the order of phase transition changes from crossover to first-order, and a CEP appears in the phase diagram. For instance, when the angular velocity increases to about 0.015 GeV, the system undergoes a transition from crossover to first-order phase transition at the chemical potential of 0.2 GeV. The CEP induced by the angular velocity is located at $(T_{CEP}, \Omega_{CEP}) = (0.240, 0.015)$ GeV.

Fig. \ref{subfig:Tmu} also reveals that the phase lines move towards the high-temperature region with increasing angular velocity, and the CEPs continuously shift towards the high chemical potential region. When the angular velocity increases from 0 to 0.05 GeV, in the $T-\mu$ phase diagram, the CEP moves from $(T_{CEP},\mu_{CEP})=(0.237,0.18)$ GeV to $(0.324,0.27)$ GeV.

Therefore, it can be concluded that the effect of angular velocity on the deconfinement phase transition is opposite compared to the chemical potential. The finite chemical potential decreases the phase transition temperature and makes the phase from first-order to crossover. With rotation, the phase transition temperature increases and allows the phase from crossover to first-order.

\section{Chiral phase transition of the 2-flavor system under rotation}
\label{sec:chiral}

In the previous sections, we have explored the phenomenon of first-order deconfinement phase transitions under rotation. The non-zero gauge field $A_\theta$ distorts the background geometry and causes the critical temperature of phase transition to increase with the rotation. This effect is crucial for a comprehensive understanding of the phase diagram of QCD matter.

This section turns to investigate the chiral phase transition. To realize chiral phase transition in DHQCD model, we introduce 2-flavor probe brane in the anisotropic background. This model provides insights into the chiral symmetry breaking and restoration in QCD. Through this approach, we can analyze the properties of chiral phase transitions affected by the rotation. Although lattice QCD indicate that in 2-flavor case, the deconfinement phase transition is a crossover under zero chemical potential. This does not affect the qualitative conclusions of our study. Therefore, this study not alter the geometric structure of the background but further explore the chiral phase transition using the probe brane within the anisotropic background.

The action of the probe brane is shown in Eq. \eqref{equ:matter}. The non-zero vacuum expectation value $X_0=\frac{\chi(z)}{2}I_{2\times2}$ of the scalar field $X$ breaks the chiral symmetry $SU(2)_L\times SU(2)_R$ to subgroup $SU(2)_V$, with identity matrix $I_{2\times2}$. According to the holographic principle, the asymptotic expansion of the scalar field $\chi$ at UV can be written as $\chi\to m_q\zeta z+...+\sigma z^3/\zeta+..$ with quark mass $m_q$, chiral condensate $\sigma$ and constant $\zeta=\sqrt{3}/2$ \cite{Cherman:2008eh}. In this paper, the quark mass is fixed to $m_q=5/1000$ GeV. As in Ref. \cite{Chen:2019rez}, the potential of Eq. \eqref{equ:matter} is considered as the following form
\begin{eqnarray}\label{equ:potantial}
   V_X(|X|,F_{MN}F^{MN})=(\frac{m_5^2}{2}+\lambda_2F^2)\chi^2+(\upsilon_4+\lambda_4F^2)\chi^4+(\upsilon_6+\lambda_6F^2)\chi^6,
\end{eqnarray}
with free parameters $\upsilon_4$, $\upsilon_6$, $\lambda_2$, $\lambda_4$, and $\lambda_6$. According to the AdS/CFT correspondence, the square of the five-dimensional mass of the field $X$ is $m_5^2=\Delta(\Delta-4)=-3$ with dimension $\Delta=3$ operator $\langle\bar{q}q\rangle$. From Eqs. \eqref{equ:matter} and \eqref{equ:potantial}, the equation of motion of scalar field can be obtained as
\begin{eqnarray}\label{equ:eom-chi}
   &-&\frac{8 e^{-2 (A_s+B)}}{f}\chi\Omega ^2 z^2\left(\lambda_2+\lambda_4\chi^2+\lambda_6\chi^4\right)+\frac{2e^{-2A_s}}{f}\chi A_t'^2z^2\left(\lambda_2+\lambda_4\chi^2+\lambda_6\chi^4\right)\nonumber\\
   &-&\frac{e^{2 A_s}}{f z^2}\chi\left(-3+4\upsilon_4\chi^2+6\upsilon_6\chi^4\right)+\left(3A_s'+\frac{f'}{f}-\Phi'-\frac{3}{z}\right)\chi'+\chi''=0,
\end{eqnarray}
where the fields of $A_s$, $f$, $B$, and $A_t$ are solved by the equations of motion Eqs. (\ref{equ:eoms-f}-\ref{equ:eoms-at}) from the gravitational background. The suitable boundary conditions need to be set for solving Eq. \eqref{equ:eom-chi}. The scalar field $\chi$ is required to satisfy the expansion $\chi(0)= m_q\zeta z+\mathcal{O}(z)=0$ at UV, and a natural boundary condition at IR. The parameters in the potential is fixed by the phase diagram of $T-\mu$ plane. As in Ref. \cite{Chen:2019rez}, these parameters are chosen as shown in Tab. \ref{tab:parachi}. With the selection of parameters, the CEP of the chiral phase transition is located in $(T_{CEP},\mu_{CEP})=(0.265,0.16)$GeV. The temperature of chiral phase transition is higher than the results of other effective models since the calculations are performed in the probe approximation at the quenched limit.

\renewcommand\arraystretch{1.5}
\begin{table}[!htbp]
	\centering
	\begin{tabular}{|p{2cm}<{\centering}|p{2cm}<{\centering}|p{2cm}<{\centering}|p{2cm}<{\centering}|p{2cm}<{\centering}|p{2cm}<{\centering}|}
	\hline
	Parameters & $\upsilon_4$ & $\upsilon_6$ & $\lambda_2$ & $\lambda_4$ & $\lambda_6$\\
	\hline
	Values & -2 & 100 & -25 & 200 & $-\frac{1000}{3}$\\
	\hline
    \end{tabular}
	\caption{\label{tab:parachi} The parameters of the potential $V_X$. They are selected from Ref. \cite{Chen:2019rez}.}
\end{table}

It should be noted that a similar approach has been used in Ref. \cite{Chen:2022mhf}, except that the scalar field $X$ describing the chiral condensate is coupled to the gauge field through the covariant derivative $D_M=\partial_M-iA_M$. Consequently, the phase transition temperature decreases with increasing rotation. However, as shown in the Ref. \cite{Chen:2019rez}, additional terms in the potential are necessary for the consistency with the QCD phase diagram of $T-\mu$ plane. Therefore, in contrast to Ref. \cite{Chen:2022mhf}, the covariant derivative is not considered in this paper and is replaced by the potential $V_X(|X|,F_{MN}F^{MN})$. Moreover, this choice brings another advantage that the scalar field $\chi$ can be considered as only depends on the fifth dimensional coordinate $z$, i.e., the near-centre approximation. Otherwise, the inclusion of covariant derivatives would leave the equation of motion of scalar field without solutions, except for the trivial one.

As can be seen from Eq. \eqref{equ:eom-chi}, the rotational and chemical potential terms in the equation are very similar, except for the opposite sign. Therefore, it can be expected that the modification of chiral condensate by rotation should be opposite to the chemical potential. That is, when the rotation is increased, the system transforms from chiral symmetry restoration to chiral symmetry breaking.

\begin{figure}[!htbp]
    \centering
    \subfigure[real rotation]{\label{subfig:sigmaT}
    \includegraphics[width=0.48\textwidth]{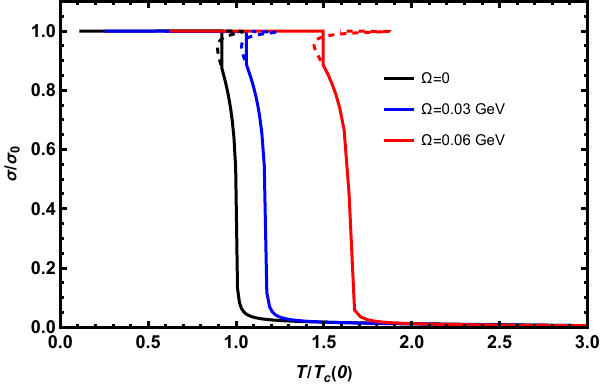}}
    \subfigure[imaginary rotation]{\label{subfig:sigmaT-i}
    \includegraphics[width=0.48\textwidth]{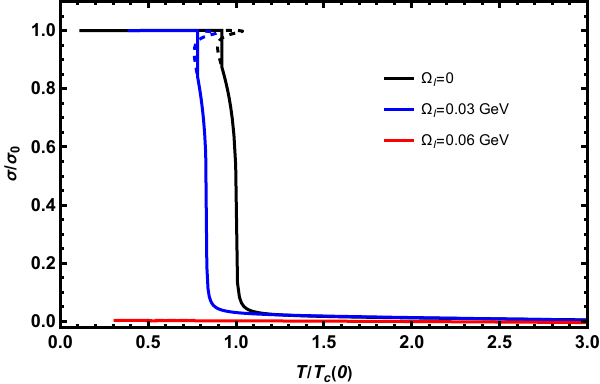}}
    \caption{\label{fig:chiral-T} The chiral condensate for 2-flavor system as a function of temperature at different angular velocities and imaginary angular velocities with chemical potential $\mu=0.12$ GeV. In the figure $\sigma_0$ is the maximum value of condensation and $T_c(0)$ is the critical temperature with zero angular velocity.}
\end{figure}

Using the equation of motion Eq. \eqref{equ:eom-chi}, we can explore the chiral condensate as a function of temperature at a chemical potential of $\mu = 0.12$ GeV for different real and imaginary angular velocities, as summarized in Fig. \ref{fig:chiral-T}. The $\sigma_0$ shown in the figure represents the value of the maximum condensate, and $T_c(0)$ indicates the critical temperature without rotation. Additionally, the dashed lines in the figure delineate the boundaries of metastable and unstable solutions.

Observing Fig. \ref{fig:chiral-T} reveals that the chiral condensate undergoes sharp changes in two distinct temperature regions, corresponding to crossover and first-order phase transitions. Specifically, only the crossover transition is considered as chiral phase transition, while the other is driven by the first-order deconfinement phase transition in the background field, which can be verified by the transition temperature and the value of condensate. Near the crossover temperature, the condensate value grows rapidly from near 0 to $0.9\sigma_0$; in the region of the first-order phase transition, the condensate value jumps from $0.9\sigma_0$ to $\sigma_0$, with a $0.1\sigma_0$ change. The first-order phase transition occurs within the temperature of deconfinement phase transition, unlike the crossover. This phenomenon further confirms that the region of significant change indeed corresponds to the true chiral phase transition.

From Fig. \ref{subfig:sigmaT}, it can be observed that the critical temperature rises with an increase in angular velocity. Fig. \ref{subfig:sigmaT-i} illustrates the suppressive effect of imaginary angular velocity on chiral condensate, which is in agreement with the findings from lattice QCD \cite{Yang:2023vsw}. It is worth noting that when the imaginary angular velocity reaches 0.06 GeV, the value of chiral condensate is quite close to 0. It can be suggested that the system is maintained in the chiral symmetry restoration phase in this case.

\begin{figure}[!htbp]
    \centering
    \subfigure[]{\label{subfig:chipdTO}
    \includegraphics[width=0.48\textwidth]{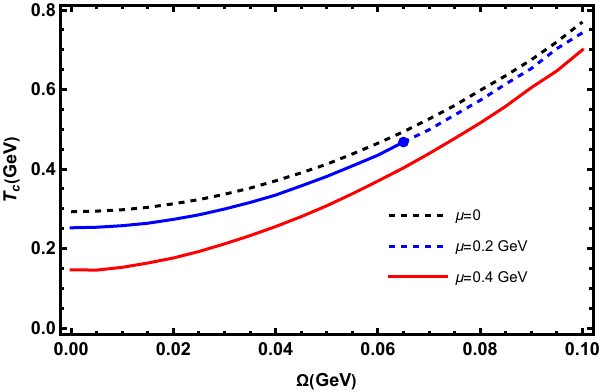}}
    \subfigure[]{\label{subfig:chipdTmu}
    \includegraphics[width=0.48\textwidth]{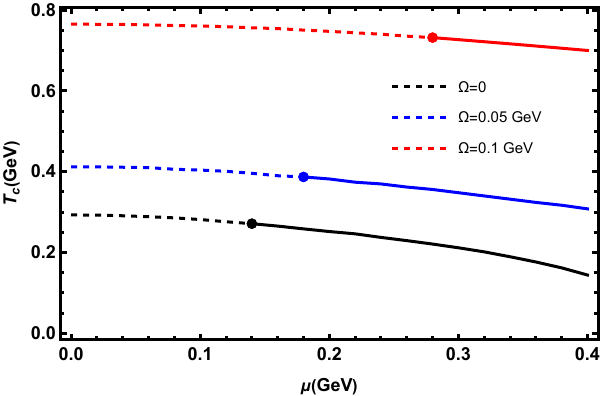}}
    \caption{\label{fig:chipd} The $T-\Omega$ and $T-\mu$ phase diagrams of chiral phase transition for 2-flavor system.}
\end{figure}

Based on the previous discussion, we can infer that the critical temperature varies with rotation. By solving Eq. \eqref{equ:eom-chi}, we have obtained the $T-\Omega$ and $T-\mu$ phase diagrams for chiral phase transition, as shown in Fig. \ref{fig:chipd}. In these diagrams, the dashed lines represent crossover, the solid lines indicate first-order phase transition, and the dots mark the location of the CEP. The temperature for crossover can be determined by the definition of the second derivative, that is, $\frac{d^2\sigma}{dT^2}=0$ or $\frac{d^2\sigma}{d\Omega^2}=0$.

From Fig. \ref{subfig:chipdTO}, it can be observed that as the angular velocity increases, the transition temperature also rises. Similar to the deconfinement phase transition, the relationship between the temperature $T_c$ and the angular velocity $\Omega$  is approximately quadratic $T_c(\Omega)/T_c(0)=1+C_2\Omega^2$. Therefore, when the angular velocity reaches 0.1 GeV, the transition temperature is about 2.6 to 4.6 times higher than that without rotation. Moreover, the finite chemical potential amplifies the coefficient $C_2$. As the angular velocity increases, the phase transition changes from first-order to crossover, and a CEP is induced in the phase diagram. Specifically, at a chemical potential of 0.2 GeV and a angular velocity of 0.065 GeV, the system transition from a first-order to crossover. In this case, the CEP appears at the position $(T_{CEP},\Omega_{CEP}) = (0.468,0.065)$ GeV.

Fig. \ref{subfig:chipdTmu} also exhibits similar behavior, where the phase boundary moves towards a higher temperature region as the angular velocity increases, and the CEP continuously moves towards the region of higher chemical potential. When the angular velocity increases to 0.1 GeV, in the $T-\mu$ phase diagram, the CEP moves to $(T_{CEP},\mu_{CEP}) = (0.732,0.28)$ GeV.

As a consequence, it can be concluded that the effect of angular velocity on the chiral phase transition is opposite to that of the chemical potential, similar to the observation in deconfinement phase transition. With finite chemical potential, the phase transition temperature is decreased and allows the phase change from crossover to first-order. As the angular velocity increases, the phase transition temperature increases and the phase crosses from first-order to crossover.

\section{Conclusion and discussion}
\label{sec:sum}

In this paper, we use a new approach to investigate the effect of rotation on deconfinement and chiral phase transitions in the dynamical holographic QCD model. Our holographic calculations are consistent with lattice QCD data \cite{Braguta:2021jgn} and \cite{Yang:2023vsw}.

By utilizing the non-trivial term $A_\theta=\Omega r^2$ in the gauge field and the dilaton field influenced by rotation, we have constructed an anisotropic gravitational background. This background can describe rotating QCD matter under the near-centre approximation. The new parameter $\mu_\Omega$ in the model is determined by the relationship between the transition temperature and the imaginary angular velocity $T/T_c \sim 1 - C_2 \Omega_I^2$ from lattice data \cite{Braguta:2021jgn}. Additionally, we also introduced a 2-flavor probe brane in this background to qualitatively analyze chiral phase transition under rotation. It is observed that at low chemical potentials, the deconfinement phase transition of pure gluon system is of first order and the chiral phase transition of 2-flavor system is of crossover. 

Using the holographic method, we calculated various thermodynamic quantities $Q$ for different angular velocities. We found that the effect of rotation on the entropy density, pressure, square of the speed of sound, and latent heat can be approximated by the relationship $Q(T, \Omega) = Q(T(1 + C_2 \Omega^2), 0)$. However, the energy density and trace anomaly do not follow this pattern. In particular, rotation smooths the energy density curve within the temperature range of $ 1-3 T_c(0) $, and the peak of the trace anomaly is suppressed as the angular velocity increases.

The deconfinement phase transition in the pure gluon system under rotation is extracted from the Polyakov loop and the chiral restoration phase transition is extracted from the chiral condensate in 2-flavor probe brane. In the DHQCD model, the effect of rotation on the two phase transitions shows consistency. Both the critical temperatures decrease/increase with imaginary/real angular velocity ($\Omega_I/\Omega$) as $T/T_c\sim 1- C_2 \Omega_I^2$ and $T/T_c\sim 1+ C_2 \Omega^2$, which is consistent with lattice QCD results. 

Furthermore, in the temperature-chemical potential $T-\mu$ phase diagrams, the critical end point (CEP) moves towards regions of higher temperature and chemical potential with real angular velocity. Specifically, when the angular velocity increases from 0 to 0.05 GeV, the CEP of the deconfinement phase transition moves from $ (T_{CEP}, \mu_{CEP}) = (0.237, 0.18) $ GeV to $ (0.324, 0.27) $ GeV, and the CEP of the chiral phase transition moves from $ (T_{CEP}, \mu_{CEP}) = (0.271, 0.14) $ GeV through $ (0.387, 0.18) $ GeV to $ (0.732, 0.28) $ GeV as the angular velocity increases from 0 to 0.1 GeV.

This study differs from previous ones \cite{Chen:2020ath} and \cite{Chen:2022mhf}. The Ref. \cite{Chen:2020ath} introduced rotation through a coordinate transformation and yielded results opposite to this study. We believe that physical results should not depend on the choice of coordinate system, so the discrepancy between the two methods needs further investigation. In the Ref. \cite{Chen:2022mhf} a different type of coupling is used, i.e. via covariant derivative $D_M=\partial_M-iA_M$, whereas $V_X(X,F^2)$ is used in this study. Different couplings can lead to different results. If both couplings are considered simultaneously, it may result in non-monotonic outcomes.

In this work, we only considered results under the near-center approximation, which is a preliminary analysis of rotation effects and does not include finite volume effect and inhomogeneity. In future work, we will consider the finite-sized rotating black brane and radial dependence in the DHQCD model to reveal more information about the rotating quark-gluon plasma and the QCD phase structure. Additionally, the 2-flavor system is treated as a probe in the current work. Therefore, the critical temperature of chiral phase transition deviates much from lattice results. In the future, the full solution including the back-reaction from the probe action to the background should be taken into account.

\begin{acknowledgments}
This work is supported by the National Natural Science Foundation of China (NSFC) Grant Nos: 12305136, 12235016, 12275108, 12221005, and the start-up funding of Hangzhou Normal University under Grant No. 4245C50223 204075, and the Strategic Priority Research Program of Chinese Academy of Sciences under Grant No XDB34030000,  and the Fundamental Research Funds for the Central Universities. Xun Chen is supported by the Research Foundation of Education Bureau of Hunan Province, China(Grant No. 21B0402) and the Natural Science Foundation of Hunan Province of China under Grants No.2022JJ40344.
\end{acknowledgments}

\bibliographystyle{unsrt}
\bibliography{ref}

\end{document}